\documentclass[sigplan,screen,nonacm]{acmart}

\usepackage{booktabs}
\usepackage{amsmath}
\usepackage{enumitem}
\usepackage{graphicx}
\usepackage{algorithm}
\usepackage{algpseudocode}
\title{Characterizing VLA Models: Identifying the Action Generation Bottleneck for Edge AI Architectures\vspace{-5pt}}

\author{Manoj Vishwanathan$^{1,2}$, Suvinay Subramanian$^1$, and Anand Raghunathan$^2$}
\affiliation{%
\institution{$^1$Google, Mountain View, CA}
\country{USA}
}
\affiliation{%
\institution{$^2$Purdue University, West Lafayette, IN}
\country{USA}
}

\begin{document}

\begin{abstract}
Vision-Language-Action (VLA) \cite{rt2} models are an emerging class of workloads critical for robotics and embodied AI at the edge. As these models scale, they demonstrate significant capability gains, yet they must be deployed locally to meet the strict latency requirements of real-time applications. This paper characterizes VLA performance on two generations of edge hardware, {\em viz.} the Nvidia Jetson Orin \cite{orin} and Thor \cite{thor} platforms. Using MolmoAct-7B \cite{molmo}, a state-of-the-art VLA model, we identify a primary execution bottleneck: up to {75\%} of end-to-end latency is consumed by the memory-bound action-generation phase. Through analytical modeling and simulations, we project the hardware requirements for scaling to 100B parameter models. We also explore the impact of high-bandwidth memory technologies and processing-in-memory (PIM) \cite{hbmpim} as promising future pathways in edge systems for embodied AI.

\end{abstract}

\maketitle
\vspace{-10pt}
\section{Problem Statement}
Vision-Language-Action (VLA) models represent a transformative shift in embodied AI, enabling robots to move from rigid pre-programmed routines to generalized semantic reasoning. As with Large Language Models (LLMs), the advancement of these models is increasingly governed by neural scaling laws; recent research suggests that robotic task performance improves at a power-law rate~\cite{neural_scaling_laws_robotics}. To achieve true general-purpose utility in complex, real-world environments and tasks, models must scale to 10–-100B parameters to effectively synthesize world knowledge with precise sensorimotor control.

This imperative for scaling, however, faces a critical hardware barrier in real-time deployment. Safe, dynamic manipulation in physical environments requires a consistent control frequency of at least 10–20 Hz. Contemporary edge accelerators are structurally ill-equipped for the sparse, memory-bound autoregressive processing required for VLA action generation. While state-of-the-art models like Gemini Robotics 1.5~\cite{gemini_robotics} have demonstrated impressive reasoning through "two-brain" architectures, these rely on offloading capable models to powerful offboard servers while maintaining only rudimentary control loops locally. Such hybrid configurations are effective for demonstrations, but they remain far from feasible in the context of low-latency, fully autonomous edge systems.

This paper provides a systematic characterization of VLA workloads to guide the next generation of edge system designs. We profile the MolmoAct-7B model on contemporary Nvidia Jetson Orin and Thor hardware platforms to establish a baseline and identify key execution bottlenecks. Building on these measurements, we leverage an in-house simulator to project the performance of future VLA models across current and hypothetical future hardware configurations. 

\section{Background and Related Work: Vision-Language-Action (VLA) Models}
\begin{figure}[t]
    \centering
    \includegraphics[width=\linewidth]{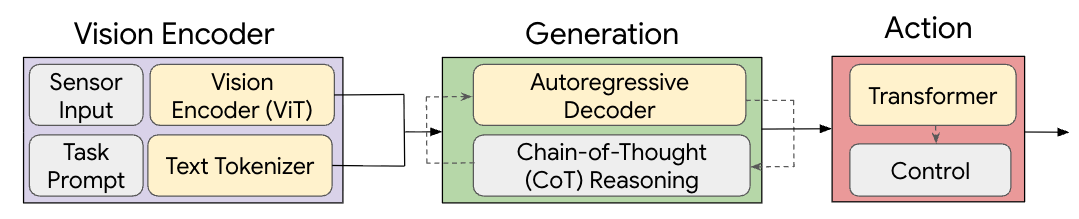}
    \Description{Block diagram of a VLA system with a vision encoder, a projector, and an autoregressive LLM backbone leading to action outputs.}
    \caption{\textbf{ VLA System Architecture. The workload consists of a Vision Encoder, a Generation Engine (autoregressive decoding), and an Action Transformer.}}
    \label{fig:vla_block}
\end{figure}
Vision-Language-Action (VLA) models are a class of multimodal foundation models that integrate visual perception, natural language understanding, and physical action. 
As illustrated in Figure~\ref{fig:vla_block} , the computational architecture of a VLA is structured into three primary subsystems:
\vspace{-3pt}
\begin{itemize}
\item Vision Encoder (Perception Core): This stage processes raw pixel data into structured feature embeddings. To capture both semantic context and spatial-geometric detail, modern VLAs often utilize fused backbones such as SigLIP and DINOv2. These high-dimensional features are mapped into the embedding space of the reasoning engine via a projector module, typically a multi-layer MLP.
\item Generation (Reasoning Engine): The core of the model is a decoder-only Transformer that processes a concatenated sequence of visual and textual tokens. During this phase, the model performs cross-modal reasoning and may generate intermediate outputs, such as "Chain-of-Thought" (CoT) reasoning or spatial waypoints, to decompose high-level instructions into executable plans.

\item Action Transformer: The final stage translates the model’s internal representations into motor commands. In discrete action tokenization, the robot’s continuous action space is quantized into bins, allowing the model to predict actions as discrete tokens within the existing vocabulary. In continuous action generation, specialized decoders such as Diffusion Transformers (DiT) are used to output smooth, high-frequency joint or end-effector trajectories.
\end{itemize}


\section{Methodology}
 Our evaluation employs a hybrid approach that combines empirical hardware profiling with high-fidelity simulated projections. This methodology allows us to establish a baseline using contemporary edge platforms and then extend those findings to evaluate the performance of larger models on hypothetical future hardware architectures.

\subsection{Real Hardware Characterization} 
To establish an empirical baseline, we characterize the performance of MolmoAct-7B on NVIDIA Jetson AGX Orin (64GB) and NVIDIA Jetson Thor (128GB) platforms. We instrument the PyTorch runtime using NVIDIA Nsight compute \cite{nsightcompute} to capture kernel-level execution traces. This profiling allows us to decompose end-to-end latency into specific phases: vision encoding, generation (autoregressive decoding), and action transformer.

\subsection{Performance Projection and Simulation
}
For modeling future architectures and larger model scales, we utilize an in-house, high-fidelity XPU simulator. This simulator incorporates detailed analytical performance models validated against several production-grade accelerators (GPUs, TPUs), achieving an accuracy of 70\% to 90\% across several production-grade models (x$--$xxx B parameter LLMs).
The simulator decomposes the VLA model into its constituent stages: vision encoding, autoregressive decoding, and action generation. Each stage is modeled as a multi-layer Transformer backbone, where each layer is further resolved into a sequence of operators, primarily high-dimensional einsums. Key features of the modeling framework include:
\begin{itemize}

\item Micro-architectural fidelity: The cost models incorporate specific hardware details, including the number of Streaming Multiprocessors (SMs), tiling strategies, and asymmetric bandwidth characteristics across different dimensions of the XPU's matrix engine.

\item Analytical roofline: The performance of individual operators is calculated using a roofline model that accounts for both compute and memory bandwidth constraints.

\item Cross-operator optimization: The framework performs optimization across operator boundaries to model effective prefetching. This is particularly critical for memory-bound operations, as it allows for early movement of operands through the memory hierarchy to minimize stalls.
\end{itemize}

\section{Evaluations and Analysis}
This section presents and analyzes the results of our measurements and simulations.

\subsection{Real Hardware Characterization}
\begin{figure}[t]
    \centering
    \includegraphics[width=\linewidth]{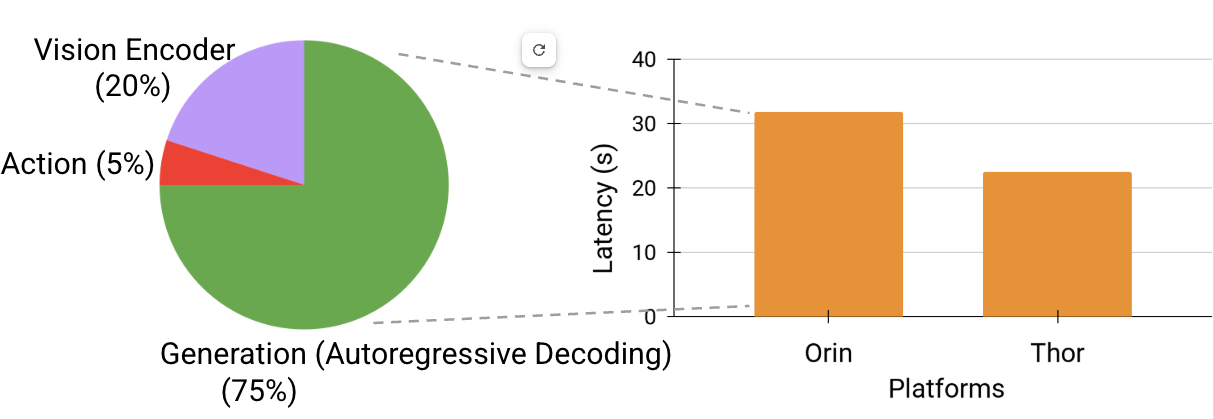}
    \Description{Performance comparison between NVIDIA Jetson Orin and NVIDIA Jetson Thor for the profiled VLA workload.}
    \caption{\textbf{Performance on current edge platforms.} Latency of MolmoAct-7B on Jetson Orin and Jetson Thor.}
    \label{fig:perf_comparison}
\end{figure}
Figure~\ref{fig:perf_comparison} summarizes our profiling results and illustrates that (i) the latencies are $\sim 200-300\times$ higher than those needed for real-time (10Hz) operation, (ii) the generation phase (auto-regressive decode with reasoning) is the primary bottleneck in VLA models, accounting for \mbox{$\approx 75\%$} of the (full-model) step latency, and
%
(iii) The generation phase is predominantly memory bandwidth bound--while Thor provides 5x the compute of Orin, the end-to-end latency only improves by 1.4x. 



\begin{table}[t]
\centering
\scriptsize
\setlength{\tabcolsep}{3pt}
\renewcommand{\arraystretch}{1.05}
\begin{tabular}{l l r r}
\toprule
\textbf{Commercial hardware} & \textbf{Memory} & \textbf{BW (GB/s)} & \textbf{BF16 TFLOPS} \\
\midrule
Orin & LPDDR5 & 203 & 100 \\
Thor & LPDDR5X & 273 & 500 \\
\bottomrule
\textbf{Hypothetical variants} & \textbf{Memory} & \textbf{BW (GB/s)} & \textbf{BF16 TFLOPS} \\
\midrule
Orin+LPDDR5X & LPDDR5X & 273 & 100 \\
Orin+GDDR7 & GDDR7 & 1000 & 100 \\
Orin+PIM & LPDDR6X PIM & 2180 & 1074 \\
Thor+GDDR7 & GDDR7 & 1000 & 500 \\
Thor+PIM & LPDDR6X PIM & 2180 & 3993 \\
\bottomrule
\end{tabular}
\vspace{5pt}
\caption{\textbf{Commercial edge platforms} and {\bf hypothetical hardware systems} used in our experiments. For systems with PIM, the compute throughput (BF16 TFLOPS) includes both the SoC and PIM.}
\label{tab:systems_commercial}
\end{table}


\vspace*{-0.1in}
\subsection{Scaled Models on Future Architectures}
\begin{figure}[h]
    \centering \includegraphics[width=1.0\linewidth]{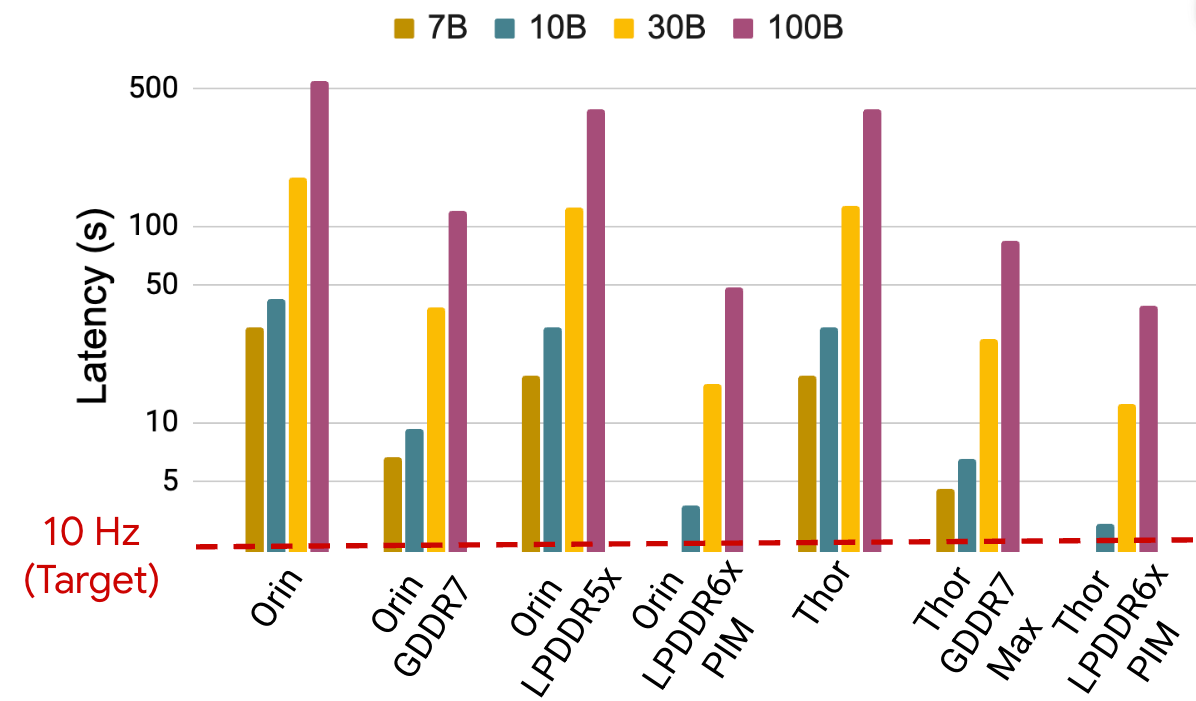}
    \Description{Bar chart of control frequency versus model size across Orin and Thor with different memory technologies, showing that higher bandwidth increases frequency but 10 Hz remains out of reach for larger models.}
    \caption{\textbf{Control Frequency for various edge system configurations} Higher memory bandwidth and PIM increase control frequency, but achieving the 10 Hz target for long horizon action generation at larger model sizes requires new innovations and algorithm-system co-design.}
    \label{fig:scaling}
\end{figure}

We scale VLA models upto 100B parameters, following scaling laws in~\cite{neural_scaling_laws_robotics,rt2}. 
We also evaluate these models in our in-house simulator on a range of current and hypothetical systems. The hypothetical systems build on a standard off-the-shelf edge hardware, but augment them with more capable memory systems (specs listed in Table~\ref{tab:systems_commercial}). 
Figure~\ref{fig:scaling} illustrates that the improved bandwidth from GDDR7 and PIM memories substantially improves performance; yet, they are
remain substantially lower than the target 10--20 Hz needed for real-world deployments.


\vspace*{-0.1in}
\section{Conclusion}
The primary bottleneck in Spatial reasoning VLA workloads is the memory-bound Action Generation phase. Standard memory scaling is insufficient for handling 10--100B parameter models at interactive rates. Future research must explore holistic system optimizations—both hardware and software to bridge the latency gap for embodied intelligence.
\section*{AI Use Statement}
The authors used Gemini (Google) to draft the Problem Statement and Related Work sections, debug PyTorch profiling scripts for the MolmoAct-7B model, and provide structural feedback on the manuscript. All AI-generated suggestions were reviewed and verified by the authors, who maintain full accountability for the technical integrity of the results.
\bibliographystyle{ACM-Reference-Format}
\bibliography{codaim_bib}
\end{document}